\documentclass[]{agujournal}

\journalname{JGR-Solid Earth}

\begin{document}

%
%

\title{Modeling Dynamic Helium Release as a Tracer of Rock Deformation.}
%

%
%

\authors{W. Payton Gardner $^{*}$\affil{1},
Stephen J. Bauer\affil{2},
Kristopher L. Kuhlman\affil{3},
and Jason E. Heath\affil{2}}

\affiliation{2}{Geomechanics Department, Sandia National Laboratories, Albuquerque, NM 87185, USA}

\affiliation{3}{Applied Systems Research Department, Sandia National Laboratories, Albuquerque, NM 87185, USA}

\affiliation{1}{Department of Geosciences, University of Montana, Missoula, MT 59812, USA}


%
%


\begin{abstract}
We use helium released during mechanical deformation of shales as a signal to explore the effects of deformation and failure on material transport properties.  A dynamic dual-permeability model with evolving pore and fracture networks is used to simulate gases released from shale during deformation and failure.  Changes in material properties required to reproduce experimentally observed gas signals are explored.  We model two different experiments of $^4$He flow rate measured from shale undergoing mechanical deformation, a core parallel to bedding and a core perpendicular to bedding.  We find that the helium signal is sensitive to fracture development and evolution as well as changes in the matrix transport properties.  We constrain the timing and effective fracture aperture, as well as the increase in matrix porosity and permeability.  Increases in matrix permeability are required to explain gas flow prior to macroscopic failure, and the short-term gas flow post failure.  Increased matrix porosity, is required to match the long-term, post-failure gas flow.  Our model provides the first quantitative interpretation of helium release as a result of mechanical deformation.  The sensitivity of this model to changes in the fracture network, as well as to matrix properties during deformation, indicates that helium release can be used as a quantitative tool to evaluate the state of stress and strain in earth materials.

\end{abstract}


\section{Introduction}
Mechanical deformation and failure of rocks can rupture mineral grains, cause pervasive microfracturing and dilation, increase effective porosity, open new fracture surfaces, and eventually cause macroscopic failure and fracture of rocks \citep{Tapponnier1976}.  These processes lead to a release of accumulated geogenic gases trapped in immobile porosity and/or mineral grains to adjacent fracture networks, which allow transport through the system (e.g. \citet{Bauer2016}).  Thus, gas release could be used as tool to monitor and investigate the state of stress and strain in earth materials and the effect of deformation on material transport properties. However, better methods for modeling and interpreting mechanically released gas need to be developed, so that this signal can be used to infer mechanical deformation and quantify the effect of deformation on gas release.

Noble gas release is indicative of the state of tectonic activity and deformation of the crust over geological time scales (e.g. \citet{Brauer2003,Kennedy2007a,Lowenstern2014}).  Noble gas isotopic composition varies with tectonic setting and crustal scale deformation \citep{Ballentine2002a}.  The crustal helium accumulation rate in regional groundwater basins is equivalent to the production rate of the whole underlying crust, which implies that pervasive fracturing and fluid migration in the lower crust are capable of releasing accumulated crustal helium and transporting it to the shallow crust \citep{Torgersen2010,Torgersen1985}.  The amount of fracturing and consequent crustal and/or mantle degassing is a function of the tectonic regime and can be correlated to tectonic velocity \citep{Kennedy2007a}.  Increased tectonism elevates crustal permeability through the formation of faults, and allows for migration of mantle helium enriched in $^3$He, and/or crustal helium enriched in $^4$He, to shallow fluid circulation systems \citep{Crossey2009}.  In areas of rapid tectonic strain and volcanic activity, radiogenic $^4$He accumulated over billions of years can be quickly released \citep{Lowenstern2014}.  

At the scale of individual strain events, a variety of geochemical and radiogenic anomalies have been observed in advance of impending seismic activity.  Radon anomalies have been reported before earthquakes (e.g. \citet{Cigolini2007,Richon2003,Wakita1980,Wakita1991}).  Radon and thoron in soil gases has been observed to fluctuate as a result of seismic activity associated with volcanic eruptions \citep{Cigolini2007,Cox1980}.  Helium isotopic ratios in springs have shown increased radiogenic $^{4}$He after earthquakes, as radiogenic helium is released from rocks during fracture and mixes into shallow groundwater systems \citep{Brauer2003}.  Radon anomalies in mine tunnels have been observed due to periodic loading from overlying hydrologic reservoir fluctuation \citep{Trique1999}. Geochemical and stable isotope anomalies have been associated with dilation and strain accumulation prior to seismicity \citep{Skelton2014,Tsunogai1995}.  Dilation and strain accumulation along faults could release radiogenic gases as a result of fracturing of mineral grains or increases in the subsurface permeability field, providing plausible mechanisms for liberating radiogenic gases prior to large scale failure.

These observations have led researchers to explore the effect of mechanical deformation on radiogenic gas emanation.  Radon emanation has been shown to follow a reproducible pattern during mechanical deformation \citep{Nicolas2014}.  Radon emanation decreases during early elastic deformation as porosity is decreased during compaction, begins to increase at about one third of the elastic yield stress due to the production of micro-fractures, and remains permanently higher after failure \citep{Holub1981,Mollo2011,Nicolas2014,Tuccimei2010}.  \citet{Bauer2016a, Bauer2016} show that accumulated helium in porespace and mineral grains is also liberated during deformation.  Radiogenic helium follows a repeatable sequence during deformation similar to radon, with decreasing helium flow rate during early compaction, increase in helium flow rate at about 1/3 of the yield strength, a sharp increase during macroscopic failure and a subsequent long term decline in gas flow.

Quantitative modeling of gas release during deformation is limited.  Recently, \citet{Girault2017} use a one-dimensional, single-domain gas transport model to investigate radon release from granites undergoing deformation during fluid pulses and triaxial deformation; however, their model does not consider matrix-fracture interactions. \citet{Holub1981} use a model of radon production and transport to simulate their observed radon signal; however, the creation of a dynamically changing fracture and/or matrix domain during and after failure was not considered. In this paper, we develop a dynamic, dual-permeability gas transport model with time dependent gas transport parameters in both the matrix and fracture domains.  This model allows us to simulate gas flow due to changing porosity and fracture aperture as a result of mechanical deformation.  We use the model and observed helium flow from shale undergoing deformation to gain insight on the deformation process and its effect on gas transport parameters.  These results set the stage for quantitative interpretation of radiogenic gas release to understand the effect of deformation and failure on gas transport properties, and the use of gas release as a means to quantitatively investigate the state of stress and strain in the earth.

\section{Theory}
Darcy's law for a compressible, non-ideal gas can be written as:
\begin{linenomath*}
\begin{equation}
	\displaystyle{
	q_{g_{x}} = \frac{Q_{g_{x}}}{A} = -\frac{k_g\bar{p}}{\mu_gzRT}\frac{\partial P}{\partial x},
	}
  \label{darcy_g}
\end{equation}
\end{linenomath*}
where $q_g$ is the molar flux (mol/s/m$^2$), $Q_g$ is the molar discharge (mol/s), $k_g$ is the permeability of the rock (m$^2$), $\mu_g$ is the gas viscosity (Pa$\cdot$s) at temperature $T$ (K), $\bar{p}$ is the average pressure in the region, $P$ is either the matrix or fracture pressure, $z$ is the dimensionless non-ideal gas factor, and $R$ is the universal gas constant (JK$^{-1}$mol$^{-1}$).  Conservation of mass gives the equation for 1D flow in the fracture as:
\begin{linenomath*}
\begin{equation}
	\displaystyle{
	\frac{\partial }{\partial t} \left(\frac{\phi_f P_f}{z}\right) = \frac{\partial}{\partial{x}}\left(\frac{k_f}{\mu_g}\frac{\bar{p_f}}{z}\frac{\partial P_f}{\partial x}\right) + R_m,
	}
  \label{flow_f}
\end{equation}
\end{linenomath*}
where $\phi_f$ is the dimensionless fracture porosity, $k_f$ is the fracture permeability and $R_m$ is an internal source (Pa/s) representing flux from the matrix to the fracture.  To simulate matrix-fracture interactions, we linearize the Darcy flux from the matrix to adjacent fracture (equation \ref{darcy_g}) over a characteristic matrix length scale to approximate the matrix contribution giving:
\begin{linenomath*}
\begin{equation}
	\displaystyle{
	R_m = -2\frac{k_mp_m}{\mu_gzb}\frac{(P_f - P_m)}{l_c},
	}
  \label{source_m}
\end{equation}
\end{linenomath*}
where the $m$ subscript denotes matrix properties, $f$ denotes fracture properties and $b$ is the fracture aperture.  Here, $l_c$ is the length of the matrix block over which the Darcy flux is linearized, assumed to be 1/2 the core diameter, which effectively treats the unfractured matrix like a single cell. The factor of two is due to the fact that there is matrix flux from both sides of the fracture.  Equation \ref{flow_f} describes gas flow in the fracture with matrix interaction.  If the matrix is considered to be at constant pressure $P_m$, then equations \ref{flow_f} and \ref{source_m} can be solved to describe the gas flux along the fracture, and the result is essentially a dual-porosity solution such as \citet{Warren1963}.

If gas discharge from the matrix at the core ends is to be considered, a dual permeability model is needed.  Compressible, real gas flow through the matrix is given by:
\begin{linenomath*}
\begin{equation}
	\displaystyle{
	\frac{\partial }{\partial t} \left(\frac{ \phi_m p_m}{z_p}\right) = \frac{\partial}{\partial{x}}\left(\frac{k_m}{\mu_g}\frac{\bar{p_m}}{z}\frac{\partial P_m}{\partial x}\right) + R_f,
	}
  \label{flow_m}
\end{equation}
\end{linenomath*}
where $\phi_m$ is the dimensionless matrix porosity and $R_f$ is the fracture interaction source-sink (Pa/s) and is approximated with the same linearization for a characteristic matrix-fracture interaction length:
\begin{linenomath*}
\begin{equation}
	\displaystyle{
	R_f = -\frac{k_m\bar{p_m}}{\mu_gz}\frac{(P_m-P_f)}{l_{c}^2}.
	}
  \label{source_f}
\end{equation}
\end{linenomath*}
In equation \ref{source_m},  the fracture cell volume per unit length of core gives the $b$ in the denominator.  In equation \ref{source_f}, the volume of matrix cell per unit length of fracture results in $l_c^2$ in the denominator.  Equations \ref{flow_f} - \ref{source_f} are a set of coupled, non-linear partial differential equations that describe gas flow through the core matrix and fracture system. These equations represent a dual-permeability system coupled by a linearized Darcy flux over a characteristic matrix block length.  

We simulate dynamic changes in fracture aperture, matrix permeability and matrix porosity by allowing $b$, $k_m$ and $\phi_m$ to change as a function of time (e.g. $b(t), k_m(t), \phi_m(t)$).  Fracture domain permeability was assigned using the fracture aperture $k_f(t) =  b(t)^2/12$ after \citet{Witherspoon1980}.  The fracture porosity ($\phi_f$) is assigned as 1, thus simulating an open fracture domain of width $b$.

\section{Methods}
\subsection{Experimental Methods}
Gas release experiments were conducted on a marine shale.  The average composition of shale samples from the same formation is 36\% clay minerals, 30\% quartz, 19\% calcite, 10\% feldspars, and 5\% other constituents.  The average porosity is around 5\%. Full details of the experimental setup and procedures can be found in \citet{Bauer2016a}.

The mechanical portion of the test system consisted of a triaxial pressure cell placed in a loading frame with the capability of testing cylindrical samples ranging in diameter from 2.5 cm (5 cm long) to 10 cm diameter (20-25 cm in length). The residual gas analysis utilized a helium leak detector, which measured the flow rate of helium.  An Oerlikon Leibold Phoenix L300i helium leak detector was used, with a minimum detectable leak rate in vacuum mode of $< 5 \times 10^{-12}$ mbar l/s. The temporal resolution of the leak rate signal was $\sim1$ s. 

Jacketed and instrumented specimens were plumbed into the base of the pressure vessel and connected to pore-pressure feed-throughs out the top of the pressure vessel. The pressure vessel was then assembled and placed into the reaction frame.  The actuator in the base of the frame was raised gradually to bring the pressure vessel piston into contact with the reaction frame.  The pressure vessel was then connected to the pressure intensifier and filled with Isopar - an incompressible, synthetic, isoparaffinic, hydocarbon confining fluid.  At this point, the servo-hydraulic control was turned on and data collection began.  The confining pressure was raised to 13.8 MPa, with the reaction frame actuator holding its position. 

\begin{figure}
 \includegraphics[width=\textwidth]{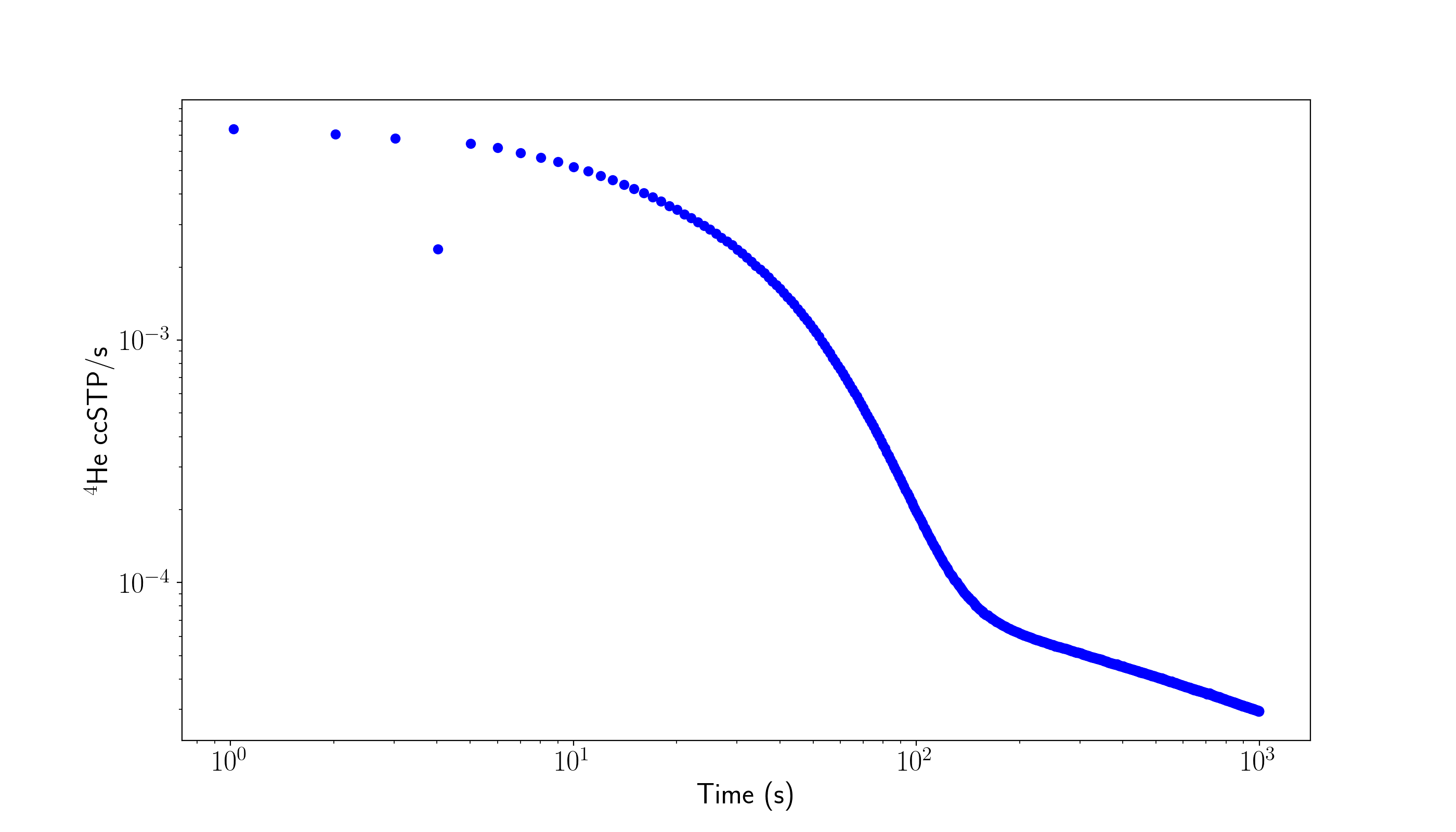}
  \caption{ SS2 experiment observed gas release for early time pump down.}
  \label{ss2_early_time}
\end{figure}

Cores were initially saturated to simulate in-situ reservoir helium concentrations by flowing helium through the sample at an inlet pressure of 0.345 MPa.  We flowed helium at a rate of around $1\times10^{-6}$ ccSTP/s for 24 hours as measured by our leak detector, adding a total volume of helium around 0.09 ccSTP, which is around 5\% of the total pore volume.  Thus, we significantly enriched our samples above the atmospheric value, calculated as $4\times10^{-6}$ ccSTP given 3\% porosity, one atmosphere of pressure and the atmospheric concentration of helium.  However, our samples are still within the range of expected natural reservoir conditions, which can be enriched over $10^5$ times above atmosphere \citep{Gardner2012}, giving a reasonable upper limit of natural helium of 0.4 ccSTP.  After saturation and prior to deformation, inlet pressure was relaxed to 0.1 MPa.  The vacuum line was then connected to the pore pressure system, allowing gas sampling access from both the top and bottom of the specimen during deformation.   During the test, vacuum was applied to both ends of the specimen.  After confining pressurization and vacuum pump down, the helium leak rate was monitored. Once the helium flow-rate reached a steady, log-linear decline, background conditions for drainage of the porous media were assumed.  The early time pump-down data prior to deformation for SS2 is shown in Figure \ref{ss2_early_time}.

Data collected in the experimental study included force, pressure, temperature, axial and lateral displacements, gas release, acoustic emissions and vacuum pressure.  In order to protect the vacuum system in the event of a confining fluid leak, low positive pressure (0.1-0.2 MPa), high-vacuum relief valves and a high-vacuum fluid expansion trap, designed to accommodate the decompression of the confining fluid from the maximum confining pressure to 0.1 MPa, were included in the vacuum system.  Helium leak rate standards were used to calibrate the leak detector.  Simulations were performed for two different gas release experiments: SS2, where the shale was cored parallel to bedding, and SS3, where the shale was cored perpendicular to bedding.  

\section{Numerical Methods}
In order to simulate gas release from core-scale deformation experiments, equations \ref{flow_f} - \ref{source_f} were simulated in a fully implicit manner using an integral finite volume method.  We developed simulation software in Python utilizing the FiPy package \citep{Guyer2009}.  Given the non-linearity of the problem, the tight coupling of the dual permeability system and abrupt transition in material properties during the macroscopic failure portion of the simulation, robust time stepping logic was implemented to facilitate a convergent solution.  Time step routines which helped in convergence include: 1) slow time step ramping, 2) continuous physically based updating of time steps using the diffusive characteristic time calculated from the dynamic fracture aperture and 3) aggressive time step cutting if the linear iterations for a non-linear time step update did not converge rapidly.

Initial and boundary conditions were applied to simulate the test conditions for each experiment.  For SS2 and SS3, the initial condition was set to $10^{5}$ Pa, which approximates the helium partial pressure after initial saturation and subsequent relaxation to ambient atmospheric pressure.  At time zero, the pressure on both ends of the core was assigned to be 0 Pa, to simulate initiation of vacuum conditions.   In order to compare model output to the experimental results recorded by the helium leak detector, the gas flow from each domain $Q_f, Q_m$ was calculated at each time step using equation \ref{darcy_g}, and the simulated pressure profile in the fracture and matrix domains at the core ends.  The area of fracture, $A_f$, is taken as the core diameter times the fracture aperture and the area of matrix was calculated from the core face area before deformation.  The total modeled gas discharge is taken as the sum of fracture and matrix discharge $Q_t = Q_f+Q_m$.

\section{Results}
Modeling was carried out for experiments SS2 and SS3.  For each experiment, initial matrix porosity and permeability were estimated by fitting the model to pre-deformation data using a Levenberg-Marquardt least squares optimization scheme.  The initial ``fracture" aperture was set such that the fracture domain had the same permeability as the matrix domain.  This effectively added a small additional matrix area during early times.  The aperture of this initial ``fracture" is on the order of 2 nm for the SS2 experiment and 1 nm for the SS3 experiment and this approximation did not measurably change the modeled gas release before fracturing.  For experiment SS2, a series of simulations explored the effect of changing matrix and fracture parameters on the model-predicted signal.  A final, manually calibrated ``preferred model" was then matched to the general characteristics of the observed gas release for the SS2 experiment.  For experiment SS3, a final ``preferred" model was created to explain the general characteristics of the observed signal and allow for comparison between SS2 and SS3 experiments.

\subsection{Exploring the Deformation Signal - SS2 Results}
\begin{figure}
 \includegraphics[width=\textwidth]{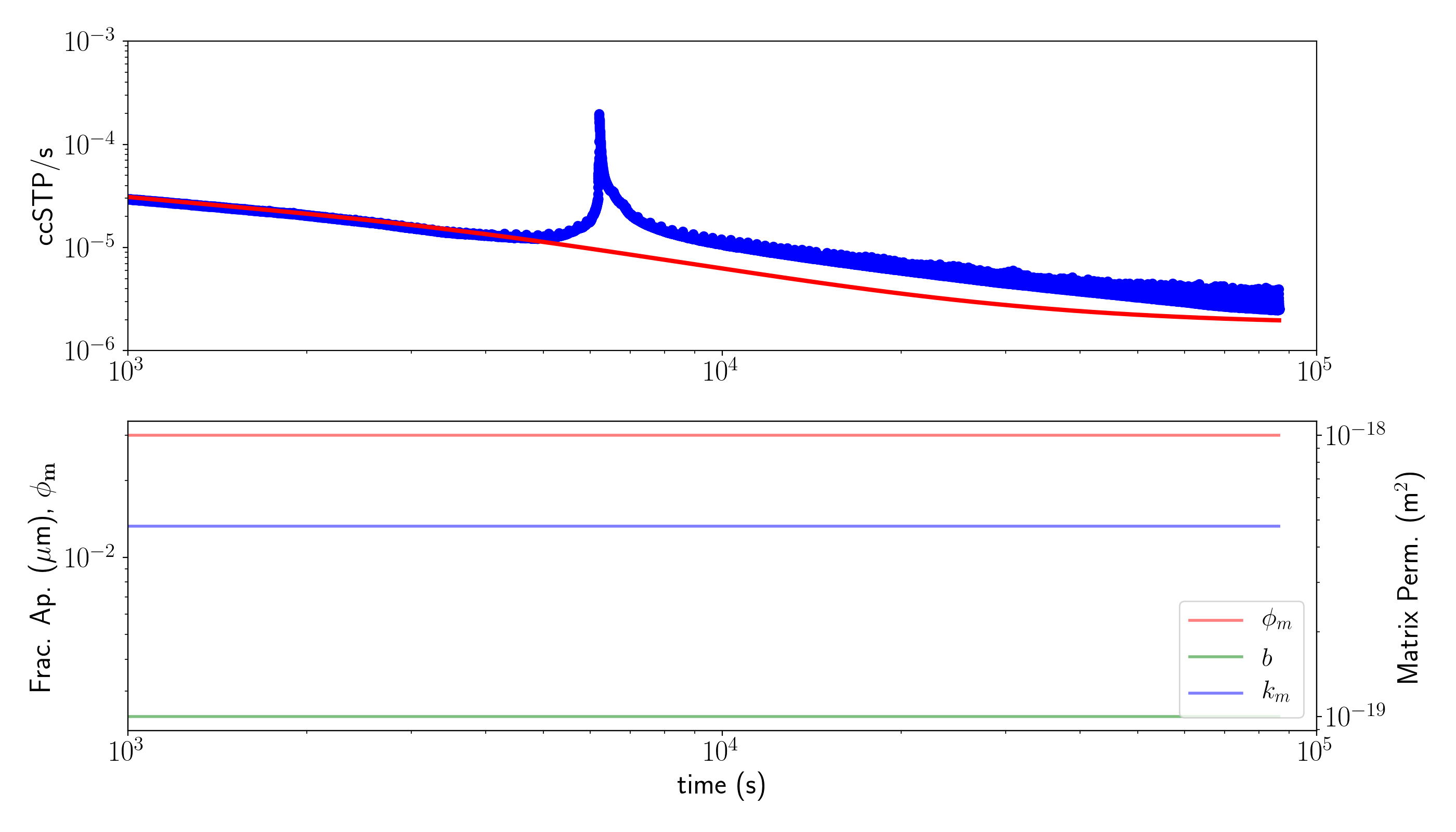}
  \caption{ SS2 experiment observed gas release shown in blue and modeled gas release shown in red (top panel) for constant, pre-deformation best fit matrix porosity and matrix permeability values (bottom).}
  \label{ss2_single_perm}
\end{figure}

The results of pre-deformation gas release modeling are compared to the observed deformation release signal in Figure \ref{ss2_single_perm}.  The estimated initial matrix permeability was $4.96\times10^{-19}$m$^2$ with a 95\% linear confidence estimate of $\pm 0.10$\%, and the estimated matrix porosity was 2.8\% with a 95\% confidence interval of $\pm 0.06\%$.  The major characteristics of the deformation release signal are described in detail in \citet{Bauer2016a}.  They include: 1) a general gas decline during the initial elastic deformation, 2) an increase in gas release due to micro-fracturing and dilation before specimen failure, 3) a sharp increase in gas release during macroscopic specimen failure, and 4) a sharp decrease in gas release immediately following fracture followed by 5) a long term slow decline in gas release for the remainder of the experiment. A single permeability model matches the early gas release well, but underestimates gas release for all times after dilation and micro-fracturing occur.  

\begin{figure}
 \includegraphics[width=\textwidth]{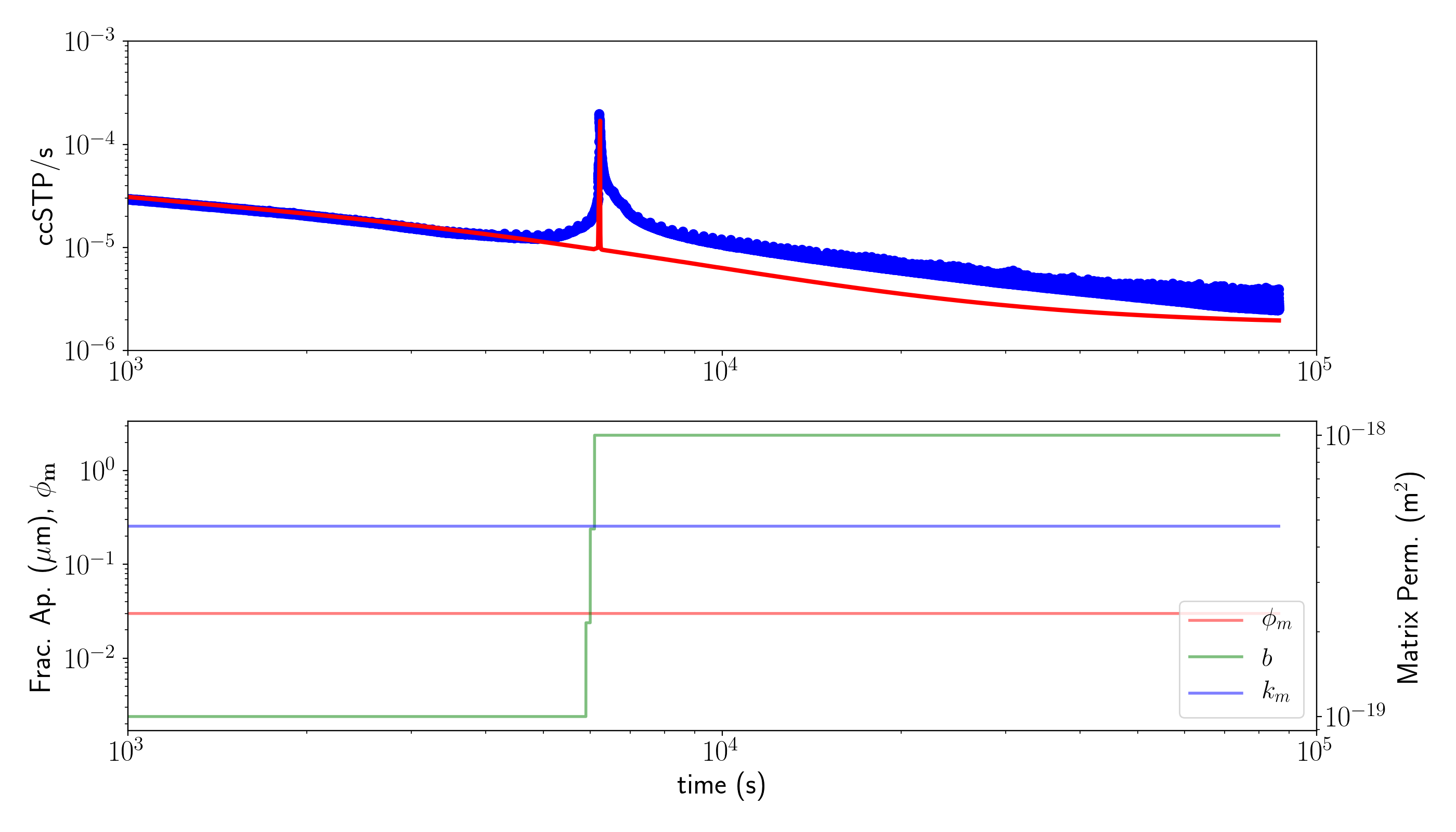}
  \caption{SS2 experiment observed gas release and modeled gas release (top) for constant, pre-deformation best fit matrix porosity and matrix permeability values and a transient fracture created at the peak of the observed gas release (bottom).}
  \label{ss2_b_variable}
\end{figure}

The simulated effects of fracture opening at the time of specimen failure are shown in Figure \ref{ss2_b_variable}.  In this model, a dynamic fracture 2 $\mu$m in aperture is opened at the time of peak gas release in the experimental signal, while all other transport parameters are kept constant.  This simulation invokes a 1000 fold increase in fracture aperture over 300 seconds.  The aperture was increased in three equal steps spaced 100 seconds apart in order to allow numerical solution of the equations during  the large-scale changes in permeability of the fracture domain.  In the simulation, gas release spikes, but the gas in the fracture domain rapidly drains and an undamaged matrix domain does not transfer sufficient gas to support the observed late time degassing signal.   This model fits the sharp peak in gas release, but does not fit pre-rupture gas release, post-rupture  gas release, or late-time gas release.  

\begin{figure}
 \includegraphics[width=\textwidth]{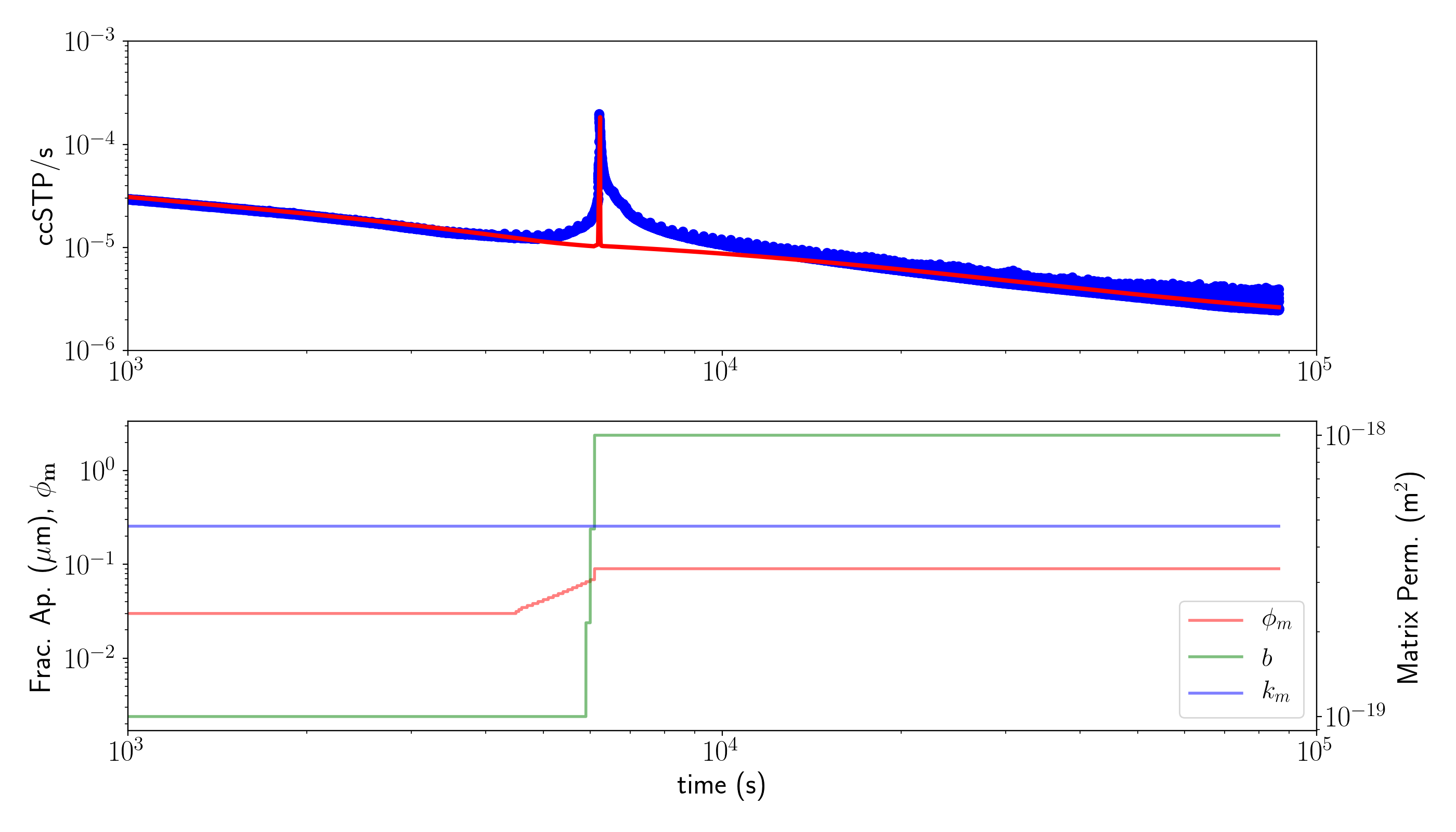}
  \caption{SS2 experiment observed gas release and modeled gas release (top) for constant matrix permeability values, transient matrix porosity and a transient fracture created at the peak of the observed gas release (bottom).}
  \label{ss2_b_phi_m_variable}
\end{figure}

Dilation and damage could increase the effective matrix porosity as pore networks become more connected by micro-fractures, thus increasing production of gas due to matrix diffusion.  Figure \ref{ss2_b_phi_m_variable} shows gas release from a model otherwise similar to Figure \ref{ss2_b_variable}, but adding increased matrix porosity during deformation.  In this simulation, we increase matrix porosity linearly from 3\% to 7\% during the pre-failure dilation period, with a subsequent jump to 9\% at the time of macroscopic failure.  These changes in volume are larger than the total volume strain or dilation observed \citep{Bauer2016}, thus the increase in matrix porosity results from increasing effective porosity due to damage.  We have no other estimates of the increase in effective porosity other than the late-time gas release signal.  The modeled increases were chosen to provide a reasonable fit to late-time gas release data.  Increasing the matrix porosity increases the effective volume available for gas release and has a large effect on the late-time degassing from the core; however, changes in the matrix porosity alone do little to effect gas release during dilation or immediately after the macroscopic failure.

\begin{figure}
 \includegraphics[width=\textwidth]{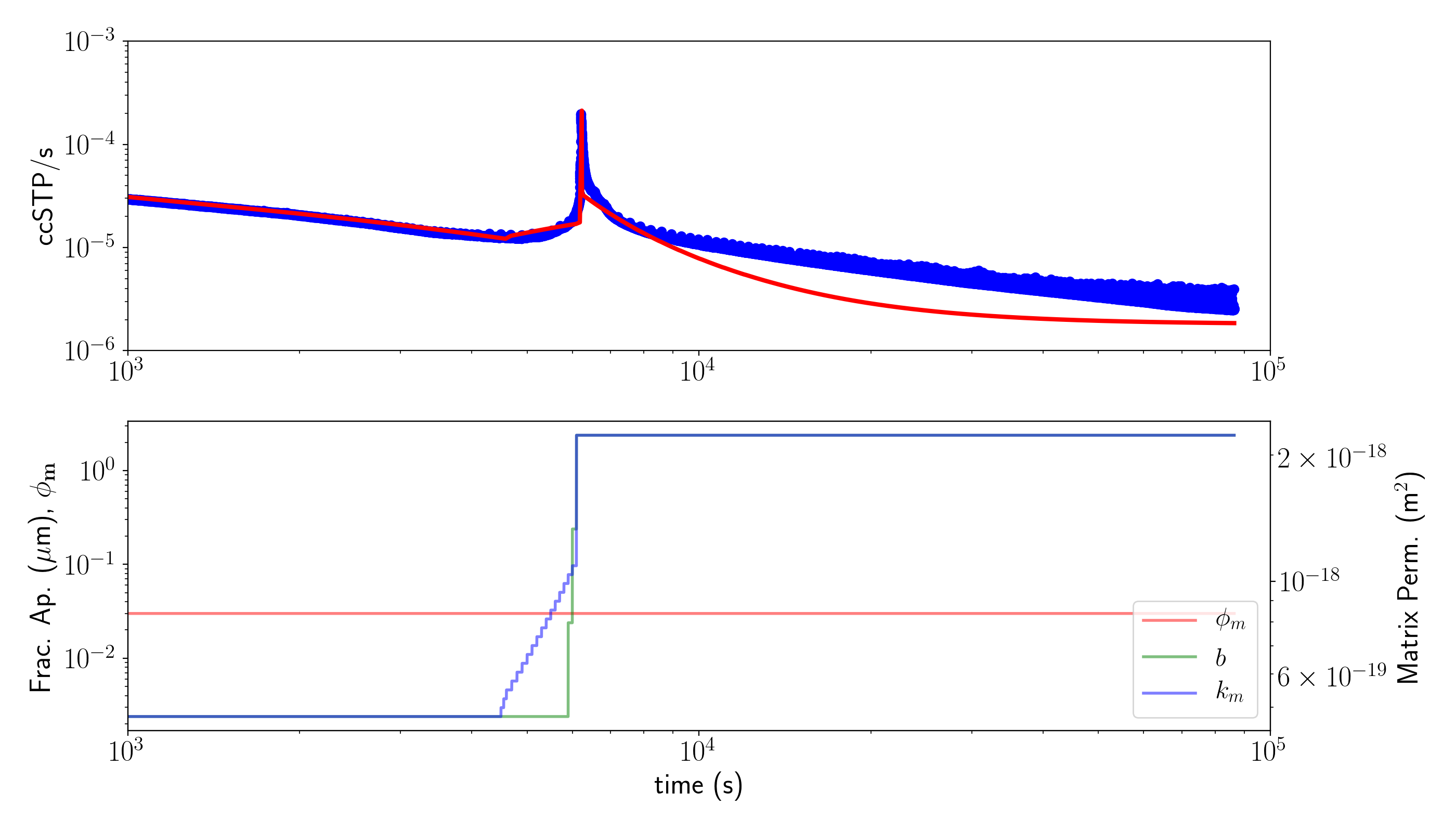}
  \caption{SS2 experiment observed gas release and modeled gas release for constant matrix porosity, transient matrix permeability and transient fracture (top panel).  Dynamic transport parameters shown in bottom panel.}
  \label{ss2_b_k_m_variable}
\end{figure}

Matrix permeability can be increased due to damage or micro-fracturing during deformation.  Figure \ref{ss2_b_k_m_variable} shows the simulated gas release for a model in which we allow fracture aperture and matrix permeability to vary during deformation but hold matrix porosity constant.  In this simulation, we linearly increase the matrix permeability from the pre-deformation value to $1\times10^{-18}$ m$^2$ during the pre-failure gas release increase.  Matrix permeability was then increased to $2\times10^{-18}$ m$^2$ at failure.  These increases were chosen to provide a reasonable match to the pre-failure and early-time, post-failure gas release.  However, increasing the matrix permeability without also increasing matrix porosity causes the core to drain too rapidly, and the late-time post-failure gas release is not well represented. 

\begin{figure}
 \includegraphics[width=\textwidth]{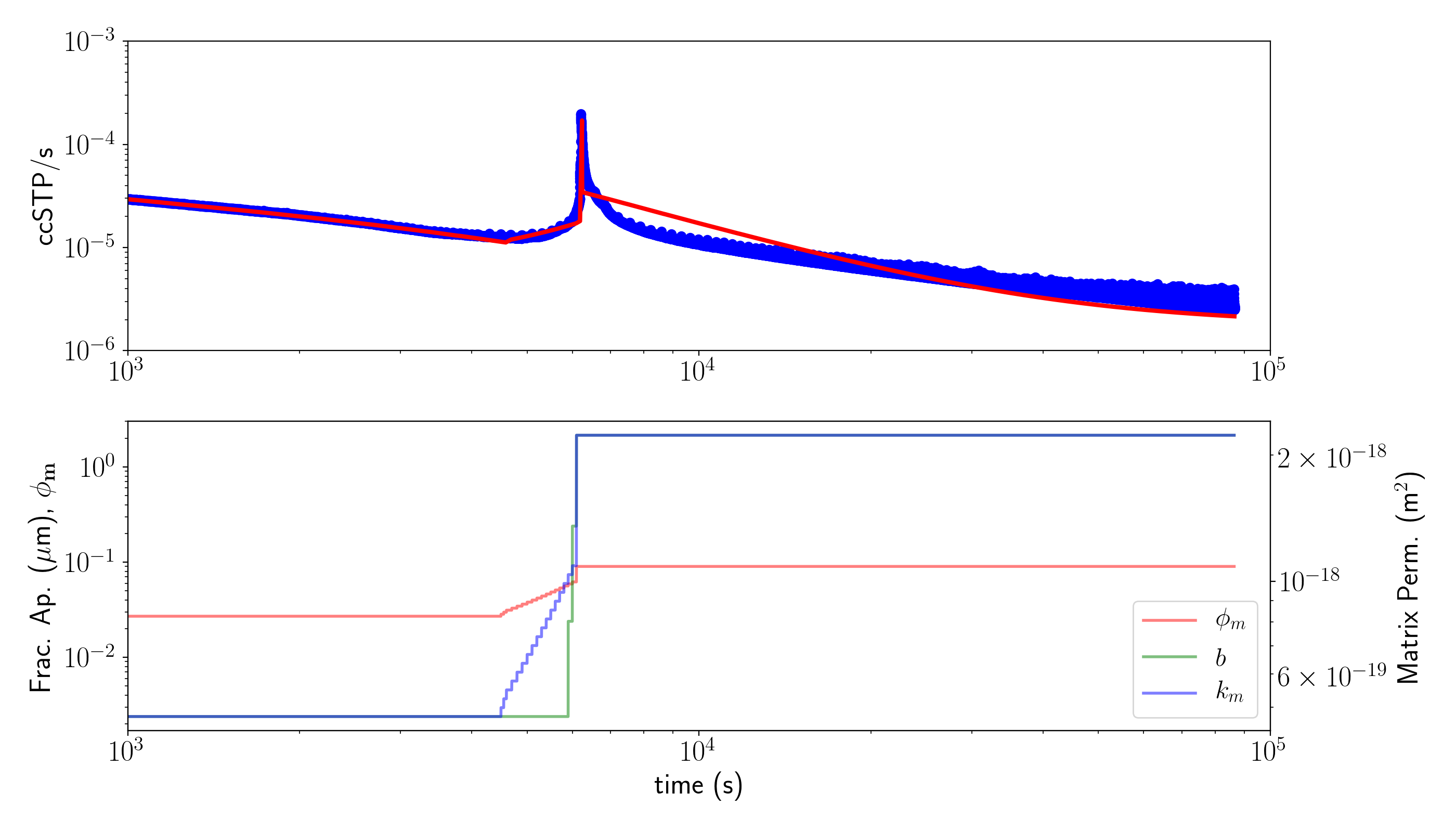}
  \caption{SS2 experiment observed gas release and modeled gas release for transient matrix porosity, matrix permeability and transient fracture (top panel).  Dynamic transport parameters are the combination of previous simulations and the time dependent values shown in bottom panel.}
  \label{ss2_b_k_m_phi_m_variable}
\end{figure}

\begin{figure}
 \includegraphics[width=\textwidth]{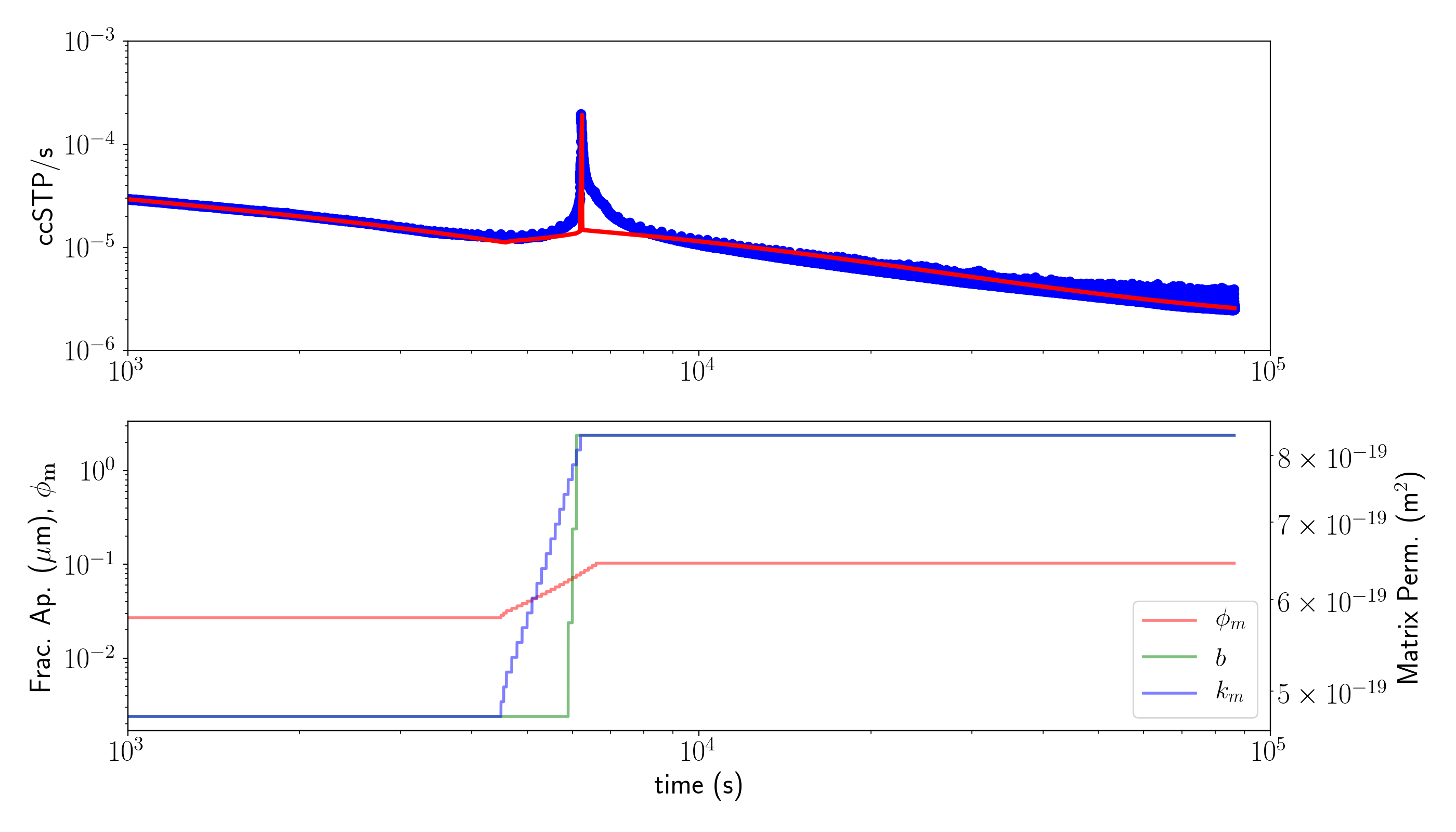}
  \caption{SS2 experiment observed and modeled gas release for a model with transient matrix porosity, matrix permeability and fracture aperture (top panel).  Dynamic transport parameters are modified from previous simulations for manual calibration to the observed signal (bottom panel.)}
  \label{ss2_b_k_m_phi_m_variable_match}
\end{figure}

We model the effects of pervasive dilation and microfracturing by combining changes in fracture aperture with the dynamic matrix porosity and permeability evolution used in the previous two simulations.  In Figure \ref{ss2_b_k_m_phi_m_variable}, we show simulated gas release for a situation in which matrix porosity, matrix permeability and fracture aperture are all variable.  In this scenario, the modeled early-time, post-fracture gas release is too high.   In Figure \ref{ss2_b_k_m_phi_m_variable_match}, we present the results of another reasonable model which reproduces the observed pre-fracture, syn-fracture and late-time, post-fracture gas release.  In this model, we reduce the increase in matrix permeability and porosity by not including a jump in these properties during macroscopic failure, and as a result fit a longer portion of the temporal gas release signal.  We have not formally calibrated our parameters, and do not claim this to be the best fit model in any mathematical sense.  The differences in the final matrix permeability and porosity between our initial fully transient model (Figure \ref{ss2_b_k_m_phi_m_variable}) and this model vary by approximately a factor of 2.

\subsection{Comparing Deformation Signals - SS3 Results}
\begin{figure}
 \includegraphics[width=\textwidth]{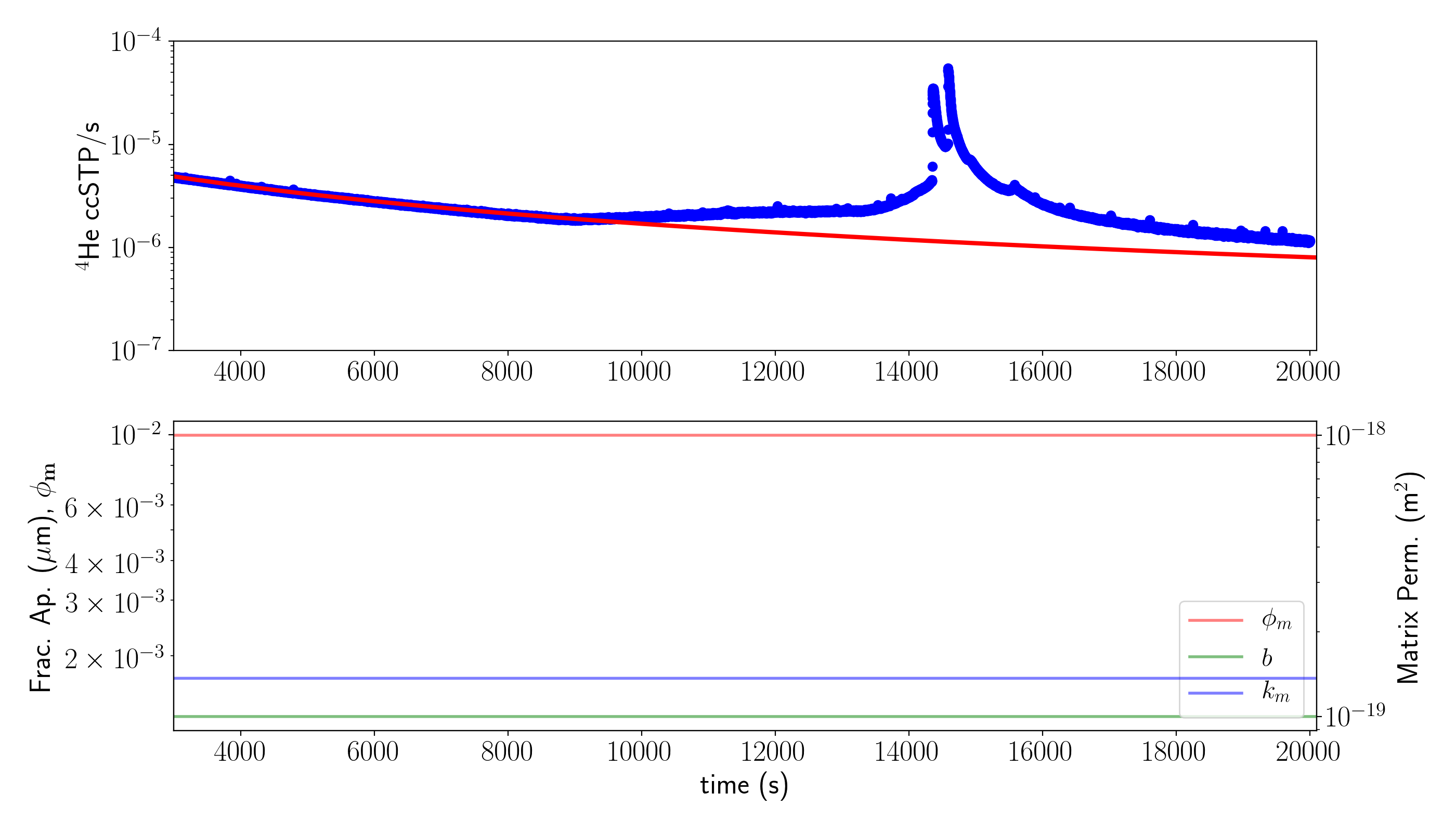}
  \caption{SS3 observed gas release and modeled gas release (top) for constant, pre-deformation best fit matrix porosity, fracture aperture and matrix permeability values (bottom).}
  \label{ss3_single_perm}
\end{figure}

In this section, we analyze the gas release signal for an experiment in which the sample was cored perpendicular to bedding.  Anisotropy in both the mechanical and transport parameters is expected in shale \citep{Metwally2011a,Gao2015}.  A constant-parameter fit to the pre-deformation data is shown in Figure \ref{ss3_single_perm}.  The estimated permeability is $1.4\times10^{-19}$ m$^2$ $\pm 0.10\%$, 3.6 times lower than when the flow is parallel to bedding, which is consistent with expected permeability anisotropy in shale \citep{Metwally2011a}.  The effective porosity perpendicular to bedding was estimated at 0.01\% which is roughly 50\% of that estimated parallel to bedding.  This sample could simply have a lower porosity than the sample used for SS2 or the connected porosity controlling flow perpendicular to bedding could be smaller than that parallel to bedding.

\begin{figure}
 \includegraphics[width=\textwidth]{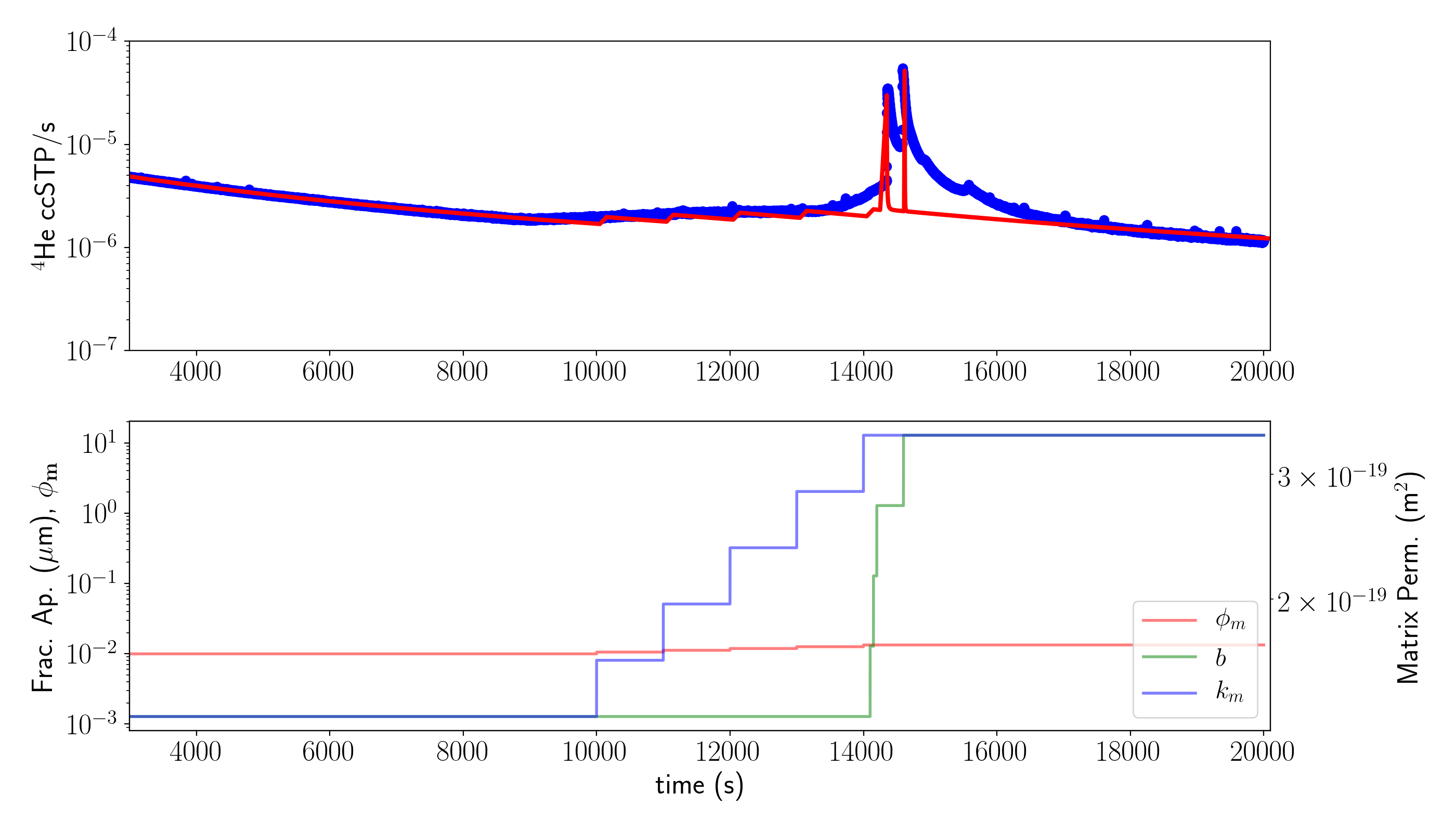}
  \caption{SS3 experiment observed and modeled gas release for transient matrix porosity, matrix permeability and fracture aperture (top panel).  Dynamic transport are modified for manual calibration to the observed signal (bottom panel).}
  \label{ss3_b_k_m_phi_m_variable_match}
\end{figure}

We use our model to investigate the differences in the gas release signal between cores parallel to (SS2) and orthogonal to (SS3) bedding.  There are significant observable differences in the gas release signal for the SS3 experiment (Figure \ref{ss3_b_k_m_phi_m_variable_match}).  Two gas release peaks are observed during macroscopic failure in the SS3 experiment, and the overall magnitude of gas release is lower in the SS3 experiment than the SS2 experiment.  During pre-failure deformation, the increase in matrix permeability required to match the pre-rupture signal in SS3 is much higher than that of the SS2 experiment.  A $50\times$ increase in matrix permeability is required to match the observed pre-release signal in SS3, bringing the matrix permeability to a value similar of that parallel to bedding. The increase in matrix porosity is of the same order as in the bedding-parallel experiment, which indicates that the overall effect of dilation in increasing effective porosity is roughly equivalent. We match the observed double peak by modeling two separate aperture increases.

\section{Discussion}
Our final models for the SS2 experiment include fracture formation, and increases in matrix permeability and porosity.  These models reproduce the general features of gas release and indicate that some amount of pervasive dilation and damage occurs throughout the core.  These results are in rough agreement with those observed by \citet{Holub1981}, who showed that radon emanation increased as micro fracturing began, and remained permanently higher after total failure of a specimen.  However, a dual permeability model cannot completely reproduce the early-time post-deformation gas release.  High matrix permeability was needed to fit the gas release signal immediately after macroscopic failure.  However, lower matrix permeability was required to fit the late-time gas release.  Thus, we hypothesize the existence of a third, limited volume, high-permeability damage zone that controls the initial post-damage response.   The remaining matrix volume, modified to a lesser extent by dilation and damage, then controls the late-time, post-failure gas release.  Examination of cores after the experiment did show the existence of smaller fractures intersecting large failure planes in both experiments, which lends qualitative support to this hypothesis \citep{Bauer2016}.

This hypothesis is also consistent with field-scale observation of damage zones, which occur near areas of high stress and strain concentration.  A near-fracture damage zone would be expected to have a higher proportion of micro fracturing and thus a higher matrix permeability.  This damage zone would drain first after failure; thus, a higher matrix permeability should be expected to control the early time gas release.  At late-time, gas release would be controlled by the less damaged matrix.  An alternative explanation is that post-failure stress relief and creep reduce the matrix permeability at long times after failure.

Another source of discrepancy between model and experimental results could be pumping of helium gas in the vacuum line itself.  Our model calculates the flow of helium (in cubic centimeters of gas at standard pressure and temperature per second (ccSTP/s)) assuming the boundary condition at the core ends are maintained at zero pressure, which is analogous to instantaneous removal of all helium released from the core ends.  In reality, the vacuum pump has a finite pumping speed, so that pressure in the vacuum line will change as a function of the gas load.  Thus, strictly speaking, our fixed pressure boundary condition is not correct.  However, the pressure in the vacuum line remained under 10$^{-3}$ mbar for the entire experiment, over 6 orders of magnitude lower than the atmospheric gas pressure in the pores.  Thus, the effect of pressure increases in the vacuum line are expected to be small.

Our results show that the helium release signal can be significantly different when the mechanical and transport properties of the material are different.   When the shale was cored perpendicular to bedding, a much larger increase in matrix permeability was required  to match observations.  This finding was consistent with post-deformation analysis of the cores which showed a single, clean fracture in the bedding-parallel core and a wider deformation zone in the bedding-perpindicular core \citep{Bauer2016a}.  

\section{Conclusions}
We developed a dynamic dual porosity and permeability model to explore gas transport from rock cores undergoing triaxial deformation.  Our model includes two permeability zones, a high permeability fracture domain and a low permeability matrix domain, both of which can undergo dynamic changes.  These domains are linked via a linear Darcy flux term which relies on the pressure distribution in both domains, thus coupling the fracture and matrix domains.  We use the Python library FiPy, an integral finite volume solver, to approximate the solution to our system of non-linear, implicitly coupled, partial differential equations.  The model is then used to reproduce gas release from an experiment where tight shale was triaxially deformed through the point of macroscopic failure.  The effect of changing transport parameters is explored, allowing us to investigate the role of deformation on gas transport in a quantitative sense.  We find that fracture creation alone does not explain the gas release, and that dilation and damage in the matrix is an important process. At early times post failure, gas release is best modeled by a increased matrix permeability, and at late-time, gas release is controlled by increased effective matrix porosity.  The helium release signal is sensitive to the amount and type of deformation occurring in a specimen. Our model can be used to probe the changes in transport parameters required to reproduce observed gas signals.  Helium signals can be used to investigate the evolution of material properties during deformation and can indicate material state during deformation.

\acknowledgments
We would like to thank Steve Ingebritsen and and an anonymous reviewer who contributed greatly to the improvement of this manuscript.  Sandia National Laboratories is a multimission laboratory managed and operated by National Technology and Engineering Solutions of Sandia, LLC., a wholly owned subsidiary of Honeywell International, Inc., for the U.S. Department of Energy's National Nuclear Security Administration under contract DE-NA0003525. We would like to thank the Sandia LDRD program for funding our initial study and equipment purchase.  The data used are listed in the references, tables, and supplemental information.  Correspondence and requests for materials should be addressed to W. Payton Gardner ~(email: payton.gardner@umontana.edu).

%

%
\bibliographystyle{agufull08}


\begin{thebibliography}{29}
\providecommand{\natexlab}[1]{#1}
\expandafter\ifx\csname urlstyle\endcsname\relax
  \providecommand{\doi}[1]{doi:\discretionary{}{}{}#1}\else
  \providecommand{\doi}{doi:\discretionary{}{}{}\begingroup
  \urlstyle{rm}\Url}\fi

\bibitem[{\textit{Ballentine et~al.}(2002)\textit{Ballentine, Burgess, and
  Marty}}]{Ballentine2002a}
Ballentine, C.~J., R.~Burgess, and B.~Marty (2002), Tracing fluid origin and
  interaction in the crust, in \textit{Noble Gases in Geochemistry and
  Cosmochemistry}, \textit{Reviews in Minerology and Geochmistry}, vol.~47,
  edited by D.~Porcelli, C.~J. Ballentine, and R.~Wieler, chap.~13, pp.
  539--614, Mineralogical Society of America.

\bibitem[{\textit{Bauer et~al.}(2016{\natexlab{a}})\textit{Bauer, Gardner, and
  Heath}}]{Bauer2016}
Bauer, S.~J., W.~P. Gardner, and J.~E. Heath (2016{\natexlab{a}}), Helium
  release during shale deformation: {E}xperimental validation,
  \textit{Geochemistry, Geophysics, Geosystems}, \textit{17}(7), 2612--2622,
  \doi{10.1002/2016GC006352}.

\bibitem[{\textit{Bauer et~al.}(2016{\natexlab{b}})\textit{Bauer, Gardner, and
  Lee}}]{Bauer2016a}
Bauer, S.~J., W.~P. Gardner, and H.~Lee (2016{\natexlab{b}}), Release of
  radiogenic noble gases as a new signal of rock deformation,
  \textit{Geophysical Research Letters}, \textit{43}(20), 10,688--10,694,
  \doi{10.1002/2016GL070876}, 2016GL070876.

\bibitem[{\textit{Brauer et~al.}(2003)\textit{Brauer, Kampf, Strauch, and
  Weise}}]{Brauer2003}
Brauer, K., H.~Kampf, G.~Strauch, and S.~M. Weise (2003), Isotopic evidence
  ({He-3/He-4, C-13(CO2)}) of fluid-triggered intraplate seismicity,
  \textit{Journal of Geophysical Research-Solid Earth}, \textit{108}(B2), times
  Cited: 37.

\bibitem[{\textit{Cigolini et~al.}(2007)\textit{Cigolini, Laiolo, and
  Coppola}}]{Cigolini2007}
Cigolini, C., M.~Laiolo, and D.~Coppola (2007), Earthquake-volcano interactions
  detected from radon degassing at {Stromboli (Italy)}, \textit{Earth and
  Planetary Science Letters}, \textit{257}(3), 511 -- 525,
  \doi{http://dx.doi.org/10.1016/j.epsl.2007.03.022}.

\bibitem[{\textit{Cox et~al.}(1980)\textit{Cox, Cuff, and Thomas}}]{Cox1980}
Cox, M.~E., K.~E. Cuff, and D.~M. Thomas (1980), Variations of ground radon
  concentrations with activity of {Kilauea Volcano, Hawaii}, \textit{Nature},
  \textit{288}(5786), 74--76.

\bibitem[{\textit{Crossey et~al.}(2009)\textit{Crossey, Karlstrom, Springer,
  Newell, Hilton, and Fischer}}]{Crossey2009}
Crossey, L.~J., K.~E. Karlstrom, A.~E. Springer, D.~Newell, D.~R. Hilton, and
  T.~Fischer (2009), {Degassing of mantle-derived CO2 and He from springs in
  the southern Colorado Plateau region -- Neotectonic connections and
  implications for groundwater systems}, \textit{Geological Society of America
  Bulletin}, \textit{121}(7-8), 1034--1053, \doi{10.1130/B26394.1}.

\bibitem[{\textit{Gao et~al.}(2015)\textit{Gao, Tao, Hu, and Yu}}]{Gao2015}
Gao, Q., J.~Tao, J.~Hu, and X.~B. Yu (2015), Laboratory study on the mechanical
  behaviors of an anisotropic shale rock, \textit{Journal of Rock Mechanics and
  Geotechnical Engineering}, \textit{7}(2), 213 -- 219,
  \doi{http://dx.doi.org/10.1016/j.jrmge.2015.03.003}.

\bibitem[{\textit{Gardner et~al.}({2012})\textit{Gardner, Harrington, and
  Smerdon}}]{Gardner2012}
Gardner, W.~P., G.~A. Harrington, and B.~D. Smerdon ({2012}), {Using excess
  $^4$He to quantify variability in aquitard leakage}, \textit{{Journal of
  Hydrology}}, \textit{{468}}, {63--75}, \doi{{10.1016/j.jhydro1.2012.08.014}}.

\bibitem[{\textit{Girault et~al.}(2017)\textit{Girault, Schubnel, and \'{E}ric
  Pili}}]{Girault2017}
Girault, F., A.~Schubnel, and \'{E}ric Pili (2017), Transient radon signals
  driven by fluid pressure pulse, micro-crack closure, and failure during
  granite deformation experiments, \textit{Earth and Planetary Science
  Letters}, \textit{474}(Supplement C), 409 -- 418,
  \doi{https://doi.org/10.1016/j.epsl.2017.07.013}.

\bibitem[{\textit{Guyer et~al.}(2009)\textit{Guyer, Wheeler, and
  Warren}}]{Guyer2009}
Guyer, J.~E., D.~Wheeler, and J.~A. Warren (2009), {FiPy}: Partial differential
  equations with {P}ython, \textit{Computing in Science \& Engineering},
  \textit{11}(3), 6--15, \doi{10.1109/MCSE.2009.52}.

\bibitem[{\textit{Holub and Brady}(1981)}]{Holub1981}
Holub, R.~F., and B.~T. Brady (1981), The effect of stress on radon emanation
  from rock, \textit{Journal of Geophysical Research: Solid Earth},
  \textit{86}(B3), 1776--1784, \doi{10.1029/JB086iB03p01776}.

\bibitem[{\textit{Kennedy and van Soest}(2007)}]{Kennedy2007a}
Kennedy, B.~M., and M.~C. van Soest (2007), Flow of mantle fluids through the
  ductile lower crust: Helium isotope trends, \textit{Science},
  \textit{318}(5855), 1433--1436, \doi{10.1126/science.1147537}.

\bibitem[{\textit{Lowenstern et~al.}(2014)\textit{Lowenstern, Evans, Bergfeld,
  and Hunt}}]{Lowenstern2014}
Lowenstern, J.~B., W.~C. Evans, D.~Bergfeld, and A.~G. Hunt (2014), Prodigious
  degassing of a billion years of accumulated radiogenic helium at
  {Y}ellowstone, \textit{Nature}, \textit{506}(7488), 355--358.

\bibitem[{\textit{Metwally and Sondergeld}({2011})}]{Metwally2011a}
Metwally, Y.~M., and C.~H. Sondergeld ({2011}), {Measuring low permeabilities
  of gas-sands and shales using a pressure transmission technique},
  \textit{INTERNATIONAL JOURNAL OF ROCK MECHANICS AND MINING SCIENCES},
  \textit{{48}}({7}), {1135--1144}, \doi{{10.1016/j.ijrmms.2011.08.004}}.

\bibitem[{\textit{Mollo et~al.}({2011})\textit{Mollo, Tuccimei, Heap,
  Vinciguerra, Soligo, Castelluccio, Scarlato, and Dingwell}}]{Mollo2011}
Mollo, S., P.~Tuccimei, M.~J. Heap, S.~Vinciguerra, M.~Soligo, M.~Castelluccio,
  P.~Scarlato, and D.~B. Dingwell ({2011}), {Increase in radon emission due to
  rock failure: An experimental study}, \textit{Geophysical Research Letters},
  \textit{{38}}, \doi{{10.1029/2011GL047962}}.

\bibitem[{\textit{Nicolas et~al.}(2014)\textit{Nicolas, Girault, Schubnel,
  Pili, Passel\`{e}gue, Fortin, and Deldicque}}]{Nicolas2014}
Nicolas, A., F.~Girault, A.~Schubnel, E.~Pili, F.~Passel\`{e}gue, J.~Fortin,
  and D.~Deldicque (2014), Radon emanation from brittle fracturing in granites
  under upper crustal conditions, \textit{Geophysical Research Letters},
  \textit{41}(15), 5436--5443, \doi{10.1002/2014GL061095}, 2014GL061095.

\bibitem[{\textit{Richon et~al.}(2003)\textit{Richon, Sabroux, Halbwachs,
  Vandemeulebrouck, Poussielgue, Tabbagh, and Punongbayan}}]{Richon2003}
Richon, P., J.-C. Sabroux, M.~Halbwachs, J.~Vandemeulebrouck, N.~Poussielgue,
  J.~Tabbagh, and R.~Punongbayan (2003), {Radon anomaly in the soil of Taal
  volcano, the Philippines: A likely precursor of the M 7.1 Mindoro earthquake
  (1994)}, \textit{Geophysical Research Letters}, \textit{30}(9),
  \doi{10.1029/2003GL016902}, 1481.

\bibitem[{\textit{Skelton et~al.}(2014)}]{Skelton2014}
Skelton, A., et~al. (2014), Changes in groundwater chemistry before two
  consecutive earthquakes in {I}celand, \textit{Nature Geosci}, \textit{7}(10),
  752--756.

\bibitem[{\textit{Tapponnier and Brace}(1976)}]{Tapponnier1976}
Tapponnier, P., and W.~Brace (1976), Development of stress-induced microcracks
  in westerly granite, \textit{International Journal of Rock Mechanics and
  Mining Sciences \& Geomechanics Abstracts}, \textit{13}(4), 103 -- 112,
  \doi{http://dx.doi.org/10.1016/0148-9062(76)91937-9}.

\bibitem[{\textit{Torgersen}(2010)}]{Torgersen2010}
Torgersen, T. (2010), Continental degassing flux of $^4${He} and its
  variability, \textit{Geochem. Geophys. Geosyst.}, \textit{11}(6), Q06,002--.

\bibitem[{\textit{Torgersen and Clarke}(1985)}]{Torgersen1985}
Torgersen, T., and W.~B. Clarke (1985), Helium accumulation in groundwater {I}:
  An evaluation of sources and the continental flux of crustal $^4${H}e in the
  {G}reat {A}rtesian {B}asin, {A}ustralia, \textit{Geochimica et Cosmochimica
  Acta}, \textit{49}, 1211--1218.

\bibitem[{\textit{Trique et~al.}(1999)\textit{Trique, Richon, Perrier, Avouac,
  and Sabroux}}]{Trique1999}
Trique, M., P.~Richon, F.~Perrier, J.~P. Avouac, and J.~C. Sabroux (1999),
  Radon emanation and electric potential variations associated with transient
  deformation near reservoir lakes, \textit{Nature}, \textit{399}(6732),
  137--141.

\bibitem[{\textit{Tsunogai and Wakita}(1995)}]{Tsunogai1995}
Tsunogai, U., and H.~Wakita (1995), Precursory chemical changes in ground
  water: {K}obe earthquake, {J}apan, \textit{Science}, \textit{269}(5220),
  61--63.

\bibitem[{\textit{Tuccimei et~al.}({2010})\textit{Tuccimei, Mollo, Vinciguerra,
  Castelluccio, and Soligo}}]{Tuccimei2010}
Tuccimei, P., S.~Mollo, S.~Vinciguerra, M.~Castelluccio, and M.~Soligo
  ({2010}), {Radon and thoron emission from lithophysae-rich tuff under
  increasing deformation: An experimental study}, \textit{Geophysical Research
  Letters}, \textit{{37}}, \doi{{10.1029/2009GL042134}}.

\bibitem[{\textit{Wakita et~al.}(1980)\textit{Wakita, Nakamura, Notsu, Noguchi,
  and Asada}}]{Wakita1980}
Wakita, H., Y.~Nakamura, K.~Notsu, M.~Noguchi, and T.~Asada (1980), Radon
  anomaly: A possible precursor of the 1978 {Izu-Oshima-Kinkai} earthquake,
  \textit{Science}, \textit{207}(4433), 882--883.

\bibitem[{\textit{Wakita et~al.}(1991)\textit{Wakita, Igarashi, and
  Notsu}}]{Wakita1991}
Wakita, H., G.~Igarashi, and K.~Notsu (1991), An anomalous radon decrease in
  groundwater prior to an {M6.0} earthquake: A possible precursor?,
  \textit{Geophysical Research Letters}, \textit{18}(4), 629--632,
  \doi{10.1029/91GL00824}.

\bibitem[{\textit{Warren and Root}(1963)}]{Warren1963}
Warren, J., and P.~Root (1963), The behavior of naturally fractured reservoirs,
  \textit{Society of Petroleum Engineers Journal}, \textit{3}(3), 245--255.

\bibitem[{\textit{Witherspoon et~al.}(1980)\textit{Witherspoon, Wang, Iwai, and
  Gale}}]{Witherspoon1980}
Witherspoon, P.~A., J.~S.~Y. Wang, K.~Iwai, and J.~E. Gale (1980), Validity of
  cubic law for fluid flow in a deformable rock fracture, \textit{Water
  Resources Research}, \textit{16}(6), 1016--1024,
  \doi{10.1029/WR016i006p01016}.

\end{thebibliography}

\clearpage

\end{document}